# HDDPM: Heteroscedastic Denoising Diffusion Probabilistic Model for Quantitative Low-Count Brain PET Recovery

Raymond Confidence[1,2] Udunna C. Anazodo[1,2]

[1]Department of Biomedical Engineering, McGill University, Montreal, Canada; [2]Montreal Neurological Institute, McGill University, Montreal, QC, Canada

**Abstract.** Positron emission tomography (PET) seeks to balance diagnostic quality with radiation dose. Low-count PET noise is non-Gaussian, non-stationary, and spatially dependent. It scales directly with local activity and is shaped by iterative reconstruction and physical corrections. Standard denoising diffusion probabilistic models (DDPMs) ignore these PET properties. Their forward process adds isotropic, homoscedastic Gaussian noise to the target. Such an approach fails to capture the realistic physical degradation generated by the imaging system. To address the above limitations, this study introduces a heteroscedastic residual diffusion model (HDDPM) for low-count brain PET recovery in which the forward corruption is itself intensity-aware. We designed a fixed, Poisson-based variance module to generate voxel-wise noise maps. These maps naturally place stronger noise perturbation on low-activity regions than high-activity ones, meanwhile the network predicts the low-to-standard-count residual under explicit dose-fraction conditioning. We evaluated our proposed model (HDDPM) alongside generative frameworks across three different scanners, using both internal and external datasets at various simulated dose levels (1% to 50%). HDDPM and isotropic DDPM showed comparable overall image quality, but HDDPM stood out in the lowest-dose (1%) external scans. It is highly reliable and significantly reduces measurement errors in both high- and low-activity regions, compared to the standard model. These results support that heteroscedastic noising with the proposed HDDPM is feasible, and it provides a physically motivated inductive bias for quantitative low-count PET recovery by reflecting the activity-dependent noise structure of PET.

## 1 Introduction

Positron emission tomography (PET) provides quantitative measurement of the metabolic and neurochemical processes required for clinical tracking of neurological disease. PET diagnostic value is directly associated with the injected radiotracer dose, which is directly proportional to image count statistics and the radiation exposure received by the patient. These competing constraints are critical in radiosensitive populations and resource-constrained settings, where the ability to reduce dose without losing quantitative image quality can mean the difference between repeat imaging and no imaging at all [1, 2]. The goal of achieving full-dose image quality from low-count acquisitions has been an active research pursuit, which aims beyond visual image quality to the exploration of PET as an intrinsic quantitative biomarker.

Low-count PET noise has three properties, all rooted in its statistical structure, and all of which compound under dose reduction. First, noise is dose-dependent: reconstructed activity error rises at low-doses, with reported root mean square error (RMSE) increasing from approximately 0.17 to 1.45 standardized uptake values (SUV; g/ml) for doses between 5% and 1% of standard dose [3]. Second, spatial non-uniformity: In uniform phantoms, local standard deviation varies up to 60% axially and 30% transaxially due to geometry-dependent attenuation and sensitivity corrections [4]. Third, activity-dependency: noise variance roughly scales with the local radioactivity concentration, consistent with Poisson statistics of coincidence detection (**Fig. 1.**) [5, 6]. Although the reconstruction literature has long established that low-count PET noise is not stationary, Gaussian, or voxel-independent [7], PET deep-learning denoising approaches have only partially incorporated this knowledge.

Most deep learning denoisers are trained to optimize image quality scores or visual appearance instead of noise structure. A neural network, such as a U-Net or residual CNN, learns a deterministic mapping at only one single implicit count level. A network that learns a deterministic map at one implicit count level rarely transfers cleanly to another. In models trained on noisier inputs, a phenomenon of over-smoothing is observed when evaluated on less nosier inputs, whilst under-denoising occurs in lower-count data [8, 9]. Recent, standard denoising diffusion probabilistic models (DDPMs) introduce a related but distinct limitation. The forward process typically corrupts the training target with spatially uniform, isotropic Gaussian noise. Although this assumption is easy to compute, it is not a good match to the reconstructed-image-domain degradation experienced in low-count PET. In this case, noise is dose-dependent, activity-dependent, spatially nonuniform, and determined by the reconstruction. Because the conventional DDPM forward variance is modulated by neither count level nor local PET signal, it motivates a heteroscedastic reformulation of the forward process for low-count PET [10, 11]. Recent studies began to address this through noise-level embeddings, dose conditioning, and unified multi-noise architectures [12, 13, 14]. Still, the forward diffusion trajectory itself has remained homoscedastic. The corruption model defined using training data does not reflect the corruption process that occurred in the scanner.

We propose a heteroscedastic diffusion probabilistic model (HDDPM) for low-count brain PET recovery, in which the forward corruption process is spatially varying and intensity-aware. A frozen count-informed variance module produces voxel-wise noise maps that assign greater perturbation to low-activity regions and lower perturbation to high-activity regions, reflecting the heteroscedastic nature of PET count statistics. The model is trained to predict the residual between low-count and standard-count PET, with dose fraction supplied as an explicit conditioning variable.

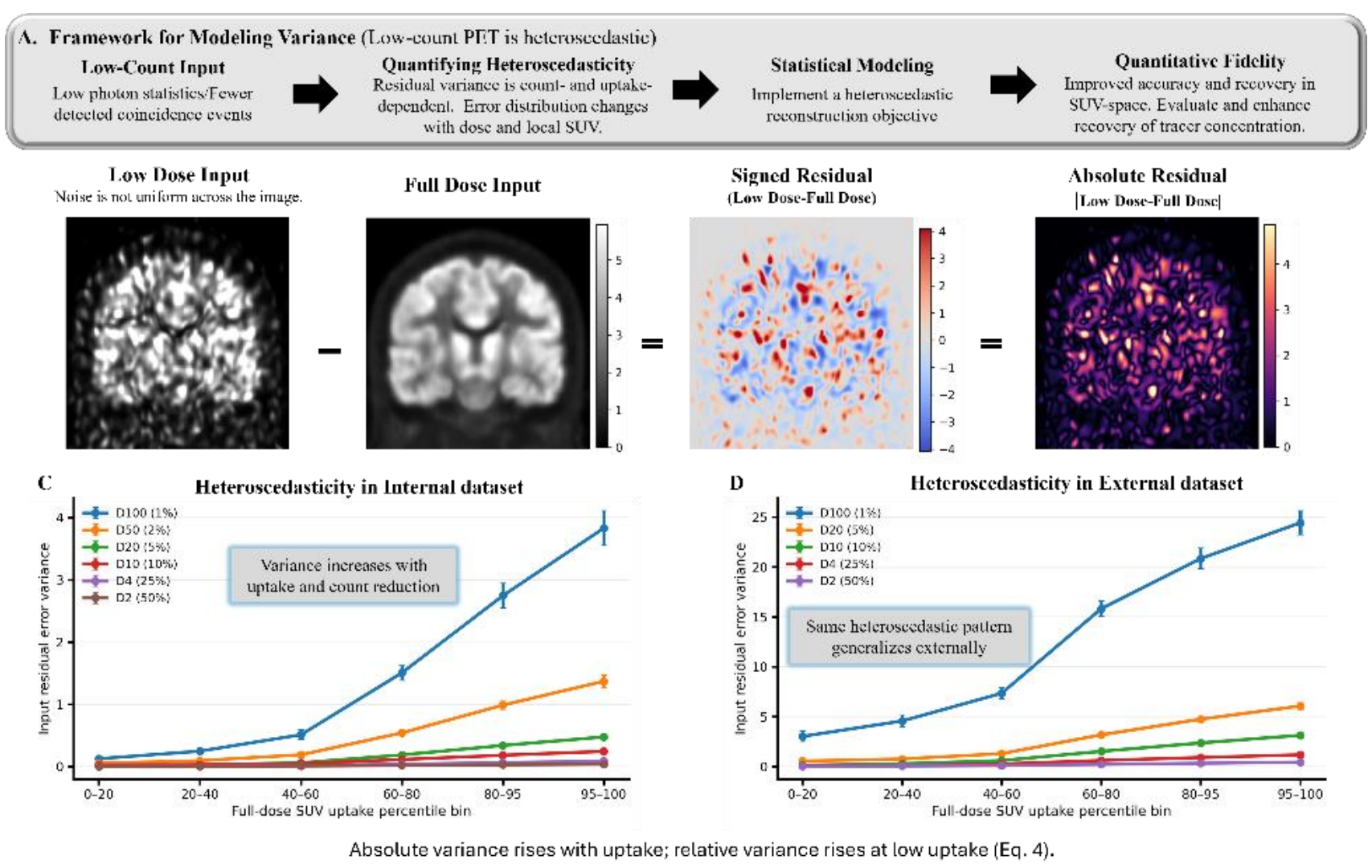

**Fig. 1.** Low-count PET heteroscedasticity motivates variance-aware diffusion.

## 2 Method

### 2.1 Problem Formulation

We frame reduced-count brain PET recovery as a paired 3D image-to-image reconstruction. For each subject, let $x_{LD} \in \mathbb{R}^{H\times W\times D}$ denote a low-count PET volume and $x_{HD} \in \mathbb{R}^{H\times W\times D}$ the corresponding standard-dose reference. The objective is to estimate $x_{HD}$ from $x_{LD}$ with preservation of anatomical structure and regional radiotracer uptake. We predict the residual rather than synthesising the target directly,

$$r = x_{HD} - x_{LD},\ \hat{x}_{HD} = x_{LD} + \hat{r}_\theta \qquad (1)$$

This residual parameterisation retains the shared anatomical and uptake structure already present in the input, redirecting the network's capacity toward the count-dependent component of the degradation. We evaluate six count-reduction levels denoted as $D_{100} = 1\%,\ D_{50} = 2\%,\ D_{20} = 5\%,\ D_{10} = 10\%,\ D_4 = 25\%,\ D_2 = 50\%$ of standard counts. Dose fraction was supplied as a conditioning variable throughout training and inference. We did not constrain diffusion timesteps to correspond to empirical dose states.

### 2.2 Heteroscedastic Residual Diffusion

A standard DDPM corrupts a clean target $r_0$ with isotropic Gaussian noise,

$$q(r_t|r_0) = \mathcal{N}\left(\sqrt{\overline{\alpha}_t}r_0,\ (1-\overline{\alpha}_t)I\right) \qquad (2)$$

where $\alpha_t$ is the cumulative noise schedule and $t$ the diffusion timestep [15]. The scalar variance is useful for analysis, but it does not fit reduced-count PET. In this scenario, local noise varies spatially with activity and depends on the count level. We replace$g(v)$ the scalar variance with a voxel-wise variance-scaling map $g(v)$, conditioned on a PET reference image $v$:

$$qHDDPMq(r_t|r_0, v) = \mathcal{N}\left(\sqrt{\overline{\alpha_t}}r_0,\ (1-\overline{\alpha_t})g(v)\right) \quad (3)$$

**Variance module.** In the configuration examined here, $g(v)$ is computed by a frozen Poisson-based module,

$$g(v) = clip\left(\frac{c}{|v|+\epsilon},\ g_{min}, g_{max}\right), \quad (4)$$

where $c$, $\epsilon$, $g_{min}$ and $g_{max}$ are fixed hyperparameters. Local relative variance is inversely proportional to local activity. This is consistent with the Poisson behaviour observed in clinical PET acquisitions [4, 5, 6]. The variance module is not learned simultaneously with the denoising network. For the main HDDPM configuration, the standard-dose target $v = x_{HD}$ serves as the variance reference during training, at inference no target-derived map is used; the trained reverse model is conditioned only on the low-count image and dose, as in the isotropic baseline. For forward sampling, the noised residual is drawn as

$$r_t = \sqrt{\overline{\alpha_t}}r_0 + \sqrt{(1-\overline{\alpha_t})g(v)\epsilon},\ \epsilon \sim \mathcal{N}(0, I) \quad (5)$$

**Denoising network.** A 3D U-Net backbone $f_\theta$ predicts the clean residual in $x_0$-prediction mode. This is represented as $\hat{r}_0 = f_\theta(r_{t,}x_{LD}, t, d, m)$, where $r_t$ is the noised residual, $x_{LD}$ is the low-count conditioning image, $t$ is the timestep embedding, $d$ is the dose fraction, and $m$ is any available metadata (including age, weight, sex, injected dose, scanner model). The recovered image is formed only after residual addition: $\hat{x}_{HD} = x_{LD} + \hat{r}_\theta$.

### 2.3 Training, Inference, and Evaluation.

**Objective.** Training minimises a residual reconstruction loss in $x_0$-prediction mode,

$$\mathcal{L}_{x_0} = ||\hat{r}_0 - r_0||\rho,$$

where $||\cdot||\rho$ denotes the configured residual regression norm. Severe low-count cases were upweighted during sampling to increase exposure to the most clinically challenging dose-reduction regimes.

**Inference.** Training was done in patches to save memory. In contrast, evaluation took place at the full-volume level. Each PET volume was divided into overlapping 3D patches, which were reconstructed patch by patch and then reassembled by overlapping aggregation. The deterministic baselines provided a single residual estimate for each patch. The DDPM and HDDPM models used deterministic 50 steps DDIM sampling [16] with a fixed sampling setup selected before the final evaluation.

**Evaluation Metrics.** We evaluated quantitative performance in raw SUV space using SUV normalized mean absolute error (SUV NMAE), SUV-based peak signal-to-noise ratio (PSNR), and structural similarity index measure (SSIM). SUV NMAE measured voxel-wise error compared to the standard-dose PET target, while PSNR and SSIM estimated global voxel-wise intensity fidelity and structural intensity preservation. For regional qualitative analysis, we computed the mean SUV, relative SUV (SUVr), signal-to-noise ratio (SNR), and contrast-to-noise ratio (CNR) in atlas-derived brain regions. The mean was the voxel-wise average within each brain region, while the SUVr was calculated for each region using cerebellar uptake as a reference. We also assessed whole-brain noise texture with noise power spectrum (NPS) summaries and evaluated high-uptake preservation with SUV p95 and p99 percent errors.

### 2.4 Implementation Details

All models were implemented in PyTorch and trained on Digital Research Alliance of Canda (Compute Canada) Narval using NVIDIA A100-SXM4 40GB GPUs. Optimization uses AdamW with a learning rate of $10^{-4}$, weight decay $10^{-4}$, and gradient clipping at 1.0. The deterministic U-Net, Pix2Pix-style GAN, residual DDPM, and heteroscedastic DDPM used a 3D U-Net backbone matched to their architecture, with $128^3$ patches and dose metadata conditioning. The U-Net and GAN baselines were trained for 129,600 updates. The GAN used an adversarial residual mapping for low-count to standard-dose. The diffusion models were trained with a batch size of 2, using residual targets defined as the difference between standard-dose and low-count PET. The residual DDPM used an $x_0$-prediction objective. In contrast, the heteroscedastic DDPM employed an analytic, spatially varying forward process along with a frozen count-informed variance module. Dose fraction was provided as an explicit conditioning variable instead of being encoded through dose-to-timestep anchoring.

## 3 Experiments and Results

### 3.1 Dataset and preprocessing

In this study, we used a part of the Ultra-low Dose PET Imaging Challenge Dataset (UDPET) [17] cohort comprised 987 unique $^{18}$F-FDG PET participants, of whom 919 were used for training and 68 reserved for validation. The training set combined 335 Siemens Biograph Vision Quadra and 584 United Imaging uExplorer participants (age 58.8 ± 13.9 years; 465 females). The full-volume evaluation was performed on 68 validation subjects (39 Siemens Biograph Vision Quadra, 29 United Imaging uExplorer; age 59.7 ± 15.6 years, 42 females; body weight 69.9 ± 19.3 kg; full injected activity 205.8 ± 46.8 MBq), evaluated at 1%, 2%, 5%, 10%, 25%, and 50% count levels. Quadra data were reconstructed using OSEM (4 iterations, 5 subsets, 2 mm Gaussian filter) with point spread function (PSF) and time-of-flight (TOF), while uExplorer data utilized 3D iterative TOF PSF with no Gaussian filter. For each subject and dose level, a paired record was constructed containing the low-count input, the standard-dose target, the residual, the dose fraction, the dose label, and any available metadata. The PET images were aligned to the MNI template with dimensions 182 x 218 x 182 using $arcsinh$ transform and 12 overlapping patches of size 128 x 128 x 128 were extracted from each PET image during training and inference. During inference, all outputs were evaluated in full-volume MNI SUV space. External validation used 67 $^{18}$F-FDG PET/MRI subjects acquired on a Siemens Biograph mMR system (age 40.2 ± 19.8 years, range 7–77; 32 females; body weight 73.5 ± 18.4 kg; estimated full injected activity 182.8 ± 34.8 MBq), evaluated at 1%, 5%, 10%, 25%, and 50% count levels [1].

### 3.2 Comparison Experiments

To compare HDDPM with deterministic regression, adversarial translation, and standard isotropic diffusion, we trained three baselines on the same paired data: a supervised 3D U-Net [18], a conditional GAN [19], and an isotropic residual DDPM [15]. To keep the comparison fair, all models were trained for the same number of steps, this means the diffusion models were not individually tuned to optimal convergence.

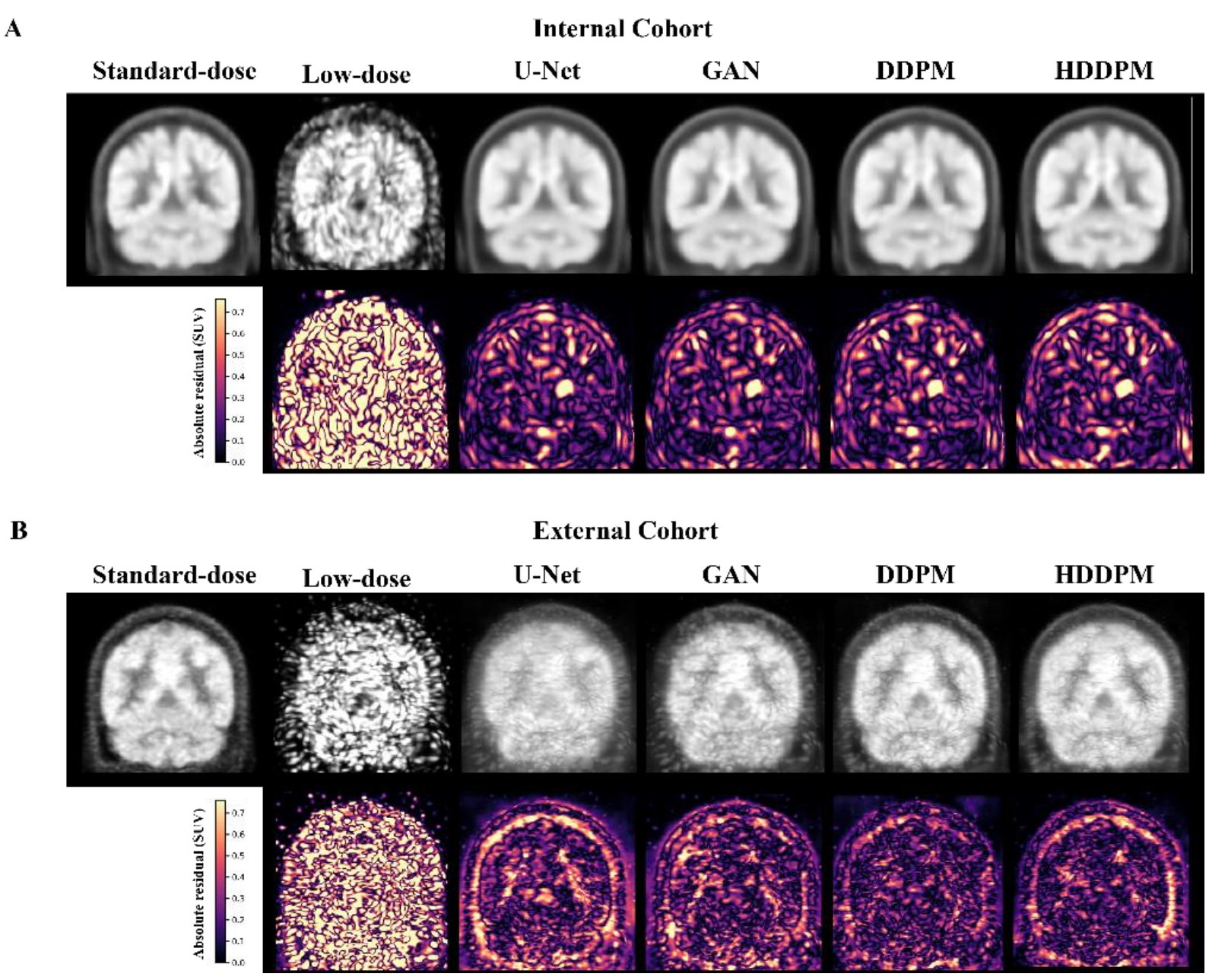

**Fig. 2.** Qualitative recovery of low-count (1%) brain PET

**Qualitative Recovery and Error Patterns.** As shown in representative internal and external PET reconstructions at the 1% effective count level **Fig. 2**. In both cases, the reconstructed images recover the standard-dose uptake pattern from the low-count input, with absolute-error maps showing reduced residual error across all models.

**Quantitative SUV Fidelity.**

**Fig. 3.** summarizes SUV-space quantitative fidelity. Across all models, performance improved significantly from 1% to 50% input counts. However, errors were consistently higher in the external cohort. Specifically, SUV NMAE decreased from 5.54-6.13% to 1.64-1.67% internally, and from 19.13-22.41% to 5.29-5.60% externally. For the diffusion models (DDPM/HDDPM), internal PSNR/SSIM rose from 30.1 dB / 0.990 to 40.6 dB / 0.999, while external values rose from 19.7 dB / 0.903 to 30.1 dB /

0.991. Absolute SUV p99 error followed a similar trend, dropping from 0.84-1.35% (internal) and 2.58-5.74% (external) at 1% to 0.14-0.38% and 0.87-2.50% at 50%.

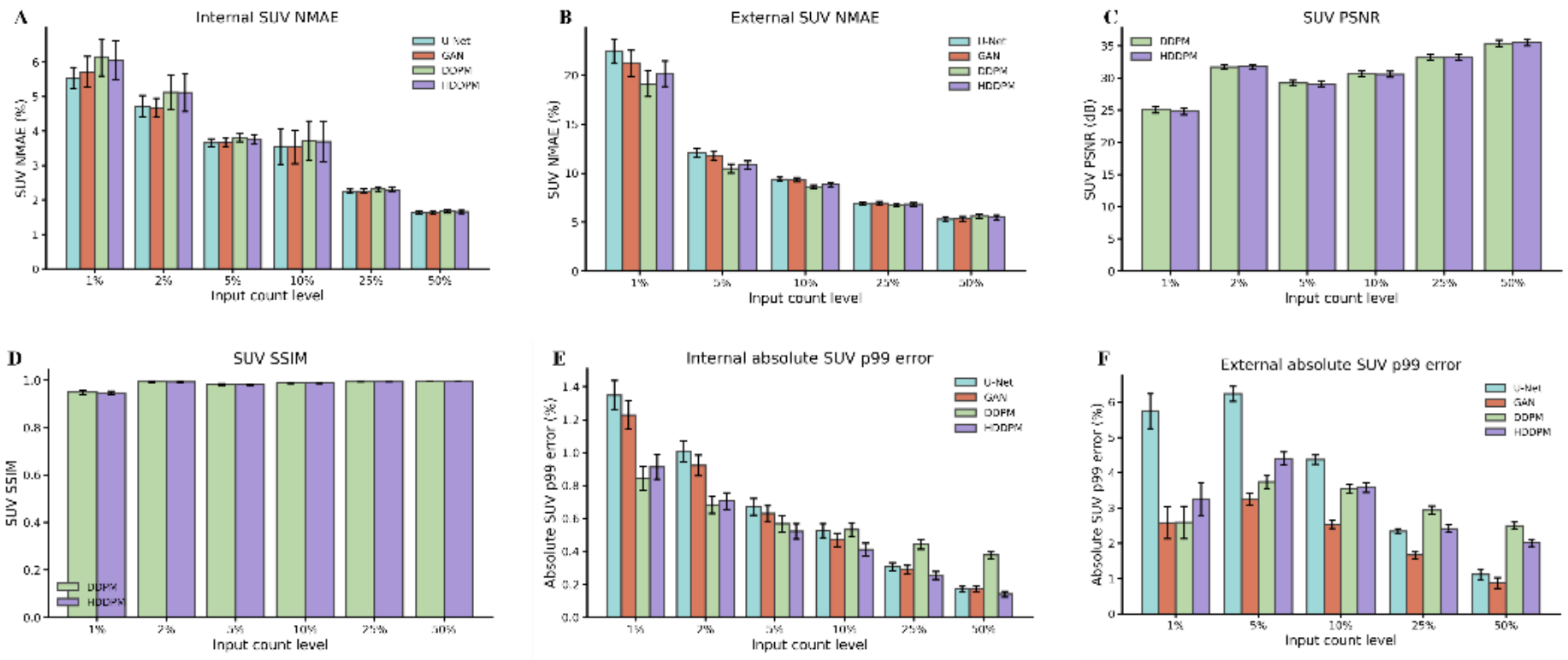

**Fig. 3.** Quantitative SUV fidelity across reduced-count levels.

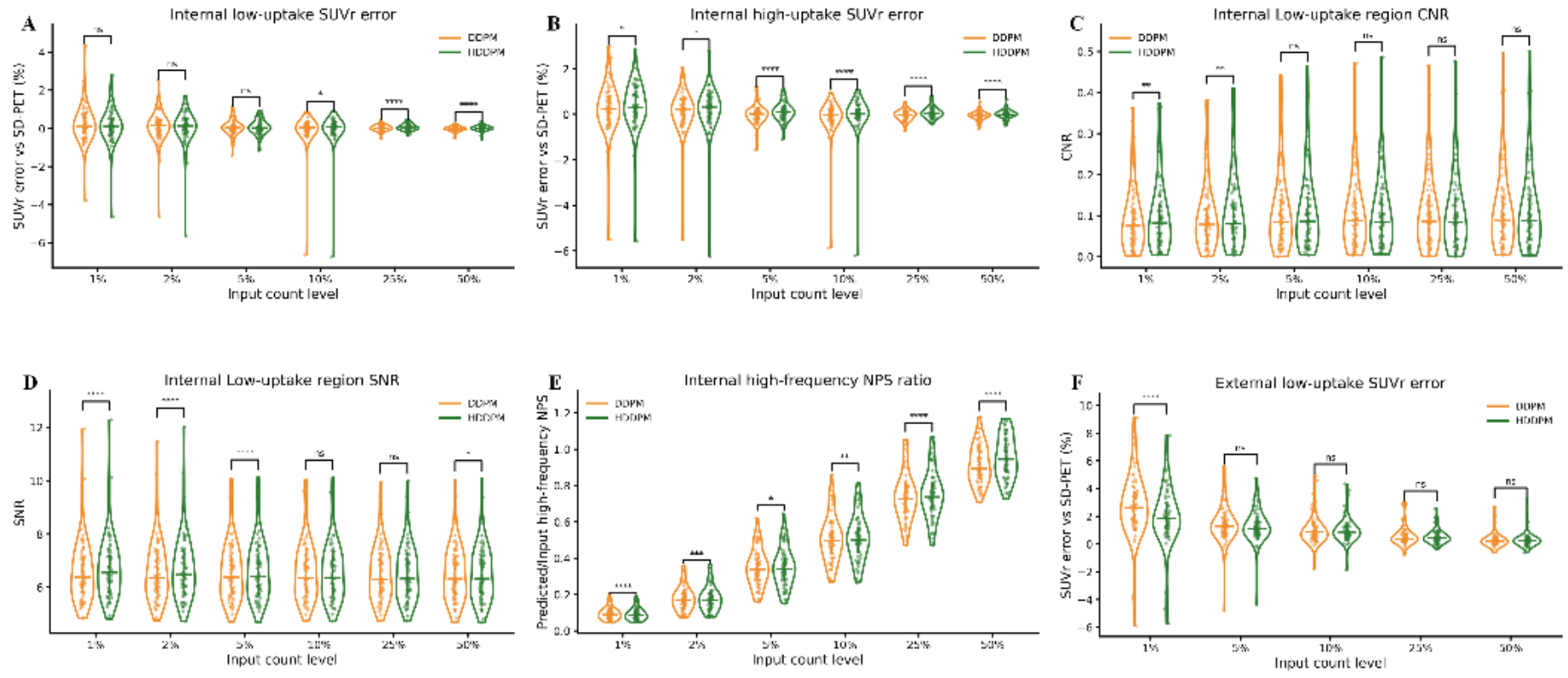

**Fig. 4.** Regional quantification and heteroscedastic diffusion analysis.

**Regional Quantification and Heteroscedastic Diffusion Analysis.** Regional quantification and noise texture comparisons reveal key differences between DDPM and HDDPM (**Fig. 3.**). In the internal cohort at 1% counts, mean absolute SUVr errors were similar between DDPM and HDDPM in both low-uptake (0.65% vs. 0.61%, p=0.49) and high-uptake regions (0.75% vs. 0.79%, p=0.206). Errors decreased uniformly to (0.09% vs. 0.10%) for low-uptake and (0.13% vs. 0.13%) for high-uptake regions at 50%. Internal noise metrics were similarly matched: low-uptake SNR was 6.58 (DDPM) vs. 6.71 (HDDPM) at 1% (6.51 vs. 6.50 at 50%), while CNR was 0.091 vs. 0.100 at 1% (0.110 for both at 50%). The high-frequency NPS power rose from 0.096 (DDPM) and 0.093 (HDDPM) at 1% to 0.914 and 0.952 at 50%. Importantly, external validation at 1% counts demonstrated a significant advantage for HDDPM, which reduced low-uptake SUVr error to 2.40% compared to DDPM's 3.10% ($p<0.001$), before both converged to (0.39% vs. 0.40%) at 50%.

## 4 Discussion

In contrast to a broad comparative study of PET denoising models, this work presents a controlled evaluation focused on assessing the value of modeling PET-specific noise properties against a strictly matched isotropic baseline. By using identical residual targets and dose conditioning for both models, this framework isolates the effect of replacing a uniform Gaussian forward process with an intensity-aware, spatially varying corruption mechanism. This distinction is needed in low-count PET, where noise is intrinsically non-uniform and inherently dependent on local activity.

Our results show that heteroscedastic diffusion provides the greatest benefit when the input is severely degraded. At the 1% count level, while HDDPM yielded only a small and non-significant decrease in internal low-uptake SUVr error compared to isotropic DDPM (0.61% vs. 0.65%), it demonstrated significantly greater robustness on external data, reducing the SUVr error from 3.10% to 2.40% ($p<0.001$). At higher count levels, the two models produced increasingly similar results. This reflects that the explicit modeling of heteroscedasticity has less effect on data with enough statistics. Additionally, comparable SUV-PSNR and SSIM values indicate that this heteroscedastic approach does not compromise overall global image fidelity.

Turning to noise texture, the NPS analysis reveals that both HDDPM and the isotropic DDPM yield similar noise-texture profiles internally, HDDPM preserves slightly greater high-frequency residual power on external data. These findings indicate that PET-dependent variance modeling improves reconstruction in challenging low-count or out-of-distribution scenarios. It does not offer the same benefits across all data types. This evidence presents HDDPM as an effective heteroscedastic extension of residual diffusion. It proves particularly useful for preserving regional quantitative accuracy in low-dose imaging.

**Limitations.** As a proof-of-concept, several architectural and training parameters for both DDPM and HDDPM, including training duration, sampler configurations, variance scaling, and model selection, were not fully optimized. Therefore, these findings should be seen as basic evidence showing the feasibility and behavior of heteroscedastic diffusion, not the best possible performance of either framework. Future work will look into a weighted-isotropic DDPM baseline to clearly separate heteroscedastic noising from low-activity reweighting.

**Conclusion.** HDDPM offers a controlled heteroscedastic extension of residual diffusion for low-count brain PET recovery. It matches the global image quality of isotropic DDPM and greatly improves regional SUVr preservation in heavily degraded or externally validated situations.

**Acknowledgments.** This study was funded by the Natural Science and Engineering Research Council (NSERC) and Fonds de recherche du Québec – Nature et technologies (FRQNT).
**Disclosure of Interests.** The authors have no competing interests to declare.